\documentclass[aps,manuscript,pop,showpacs,preprint,superscriptaddress]{revtex4-1}
\usepackage{lmodern}

\usepackage{lmodern}
\usepackage[T1]{fontenc}
\usepackage[latin9]{inputenc}
\usepackage{color}
\usepackage{amsmath}
\usepackage{graphicx}

\makeatletter

\providecommand{\tabularnewline}{\\}

\@ifundefined{textcolor}{}
{%
 \definecolor{BLACK}{gray}{0}
 \definecolor{WHITE}{gray}{1}
 \definecolor{RED}{rgb}{1,0,0}
 \definecolor{GREEN}{rgb}{0,1,0}
 \definecolor{BLUE}{rgb}{0,0,1}
 \definecolor{CYAN}{cmyk}{1,0,0,0}
 \definecolor{MAGENTA}{cmyk}{0,1,0,0}
 \definecolor{YELLOW}{cmyk}{0,0,1,0}
}

\usepackage{amsthm}\usepackage{latexsym}\usepackage{bm}\usepackage{amsfonts}

\makeatother

\begin{document}

\title{Explicit symplectic algorithms based on generating functions for
relativistic charged particle dynamics in time-dependent electromagnetic
field}

\author{Ruili Zhang}

\affiliation{Department of Modern Physics and School of Nuclear Science and Technology,
University of Science and Technology of China, Hefei, Anhui 230026,
China}

\affiliation{Key Laboratory of Geospace Environment, CAS, Hefei, Anhui 230026,
China}

\author{Yulei Wang}

\affiliation{Department of Modern Physics and School of Nuclear Science and Technology,
University of Science and Technology of China, Hefei, Anhui 230026,
China}

\affiliation{Key Laboratory of Geospace Environment, CAS, Hefei, Anhui 230026,
China}

\author{Yang He}

\affiliation{Department of Modern Physics and School of Nuclear Science and Technology,
University of Science and Technology of China, Hefei, Anhui 230026,
China}

\affiliation{Key Laboratory of Geospace Environment, CAS, Hefei, Anhui 230026,
China}

\author{Jianyuan Xiao}

\affiliation{Department of Modern Physics and School of Nuclear Science and Technology,
University of Science and Technology of China, Hefei, Anhui 230026,
China}

\affiliation{Key Laboratory of Geospace Environment, CAS, Hefei, Anhui 230026,
China}

\author{Jian Liu}

\affiliation{Department of Modern Physics and School of Nuclear Science and Technology,
University of Science and Technology of China, Hefei, Anhui 230026,
China}

\affiliation{Key Laboratory of Geospace Environment, CAS, Hefei, Anhui 230026,
China}

\author{Hong Qin }

\thanks{Corresponding author. hongqin@ustc.edu.cn}

\affiliation{Department of Modern Physics and School of Nuclear Science and Technology,
University of Science and Technology of China, Hefei, Anhui 230026,
China}

\affiliation{Plasma Physics Laboratory, Princeton University, Princeton, NJ 08543,
USA}

\author{Yifa Tang}

\affiliation{LSEC, Academy of Mathematics and Systems Science, Chinese Academy
of Sciences, Beijing 100190, China}
\begin{abstract}
Relativistic dynamics of a charged particle in time-dependent electromagnetic
fields has theoretical significance and a wide range of applications.
It is often multi-scale and requires accurate long-term numerical
simulations using symplectic integrators. For modern large-scale particle
simulations in complex, time-dependent electromagnetic field, explicit
symplectic algorithms are much more preferable. In this paper, we
treat the relativistic dynamics of a particle as a Hamiltonian system
on the cotangent space of the space-time, and construct for the first
time explicit symplectic algorithms for relativistic charged particles
of order 2 and 3 using the sum-split technique and generating functions.
\end{abstract}
\maketitle

\section{Introduction}

Dynamics of relativistic charged particles in time-dependent electromagnetic
fields has theoretical significance and a wide range of applications
in astrophysics, plasma physics, accelerator physics, quantum physics,
and many other sub-fields of physics. It often involves multi-scale
processes and long-term simulations, and geometric numerical algorithms
are required for better efficiency, accuracy and conservativeness.
Recently, advanced geometric numerical algorithms have been developed
for charged particle dynamics \textcolor{black}{\cite{qin2008variational,qin2009variational,guan2010phase,squire2012gauge,qin2013boris,liu2014fate,zhang2014,zhang2015volume,zhangapplication,ellison2015comment,he2015volume,he2015explicit,ellison2015development,liu2015neoclassical,He16-172}}
and infinite dimensional particle-field systems \textcolor{black}{\cite{Squire4748,squire2012geometric,xiao2013variational,kraus2013variational,evstatiev2013variational,zhou2014variational,Shadwick14,xiao2015variational,xiao2015explicit,crouseilles2015hamiltonian,qin2015comment,he2015Hamiltonian,qin2016canonical,zhou2015formation,Webb16}}.

Relativistic charged particle dynamics is described as a Hamiltonian
system in the canonical coordinates $(\mathbf{x},\mathbf{p})$,
\begin{equation}
\dfrac{d\mathbf{Z}}{dt}=J^{-1}\nabla H(\mathbf{Z},t):=\begin{cases}
\dfrac{d\mathbf{x}}{dt} & =\dfrac{\partial H}{\partial\mathbf{p}}=\dfrac{\left[\mathbf{p}-q\mathbf{A}(\mathbf{x},t)\right]}{\gamma m}\,,\\
\dfrac{d\mathbf{p}}{dt} & =-\dfrac{\partial H}{\partial\mathbf{x}}=-q\nabla\phi(\mathbf{x})+\left(\dfrac{\partial\mathbf{A}(\mathbf{x},t)}{\partial\mathbf{x}}\right)^{T}\dfrac{q\left[\mathbf{p}-q\mathbf{A}(\mathbf{x},t)\right]}{\gamma m}\,,
\end{cases}\label{eq:nr}
\end{equation}
where $\mathbf{Z}=(\mathbf{x}^{T},\mathbf{p}^{T})^{T}$ is a 6-dimensional
vector, $\gamma=\sqrt{1+\left[\mathbf{p}-q\mathbf{A}(\mathbf{x},t)\right]^{2}/m^{2}c^{2}}$,
\[
\thinspace J=\left(\begin{array}{cc}
0 & -I\\
I & 0
\end{array}\right)\thinspace
\]
is the canonical symplectic matrix and
\begin{equation}
\begin{split}H(\mathbf{Z},t) & \equiv\gamma mc^{2}+q\phi(\mathbf{x},t)\end{split}
\label{eq:H}
\end{equation}
is the Hamiltonian function. The canonical symplectic structure $d\mathbf{p}\wedge d\mathbf{x}$
of the exact flow of Eq.\,\eqref{eq:nr} is conserved,

\begin{equation}
\dfrac{d}{dt}(d\mathbf{p}\wedge d\mathbf{x})=d\dot{\mathbf{p}}\wedge d\mathbf{x}+d\mathbf{p}\wedge d\dot{\mathbf{x}}=0\thinspace.
\end{equation}
For canonical Hamiltonian systems equipped with canonical symplectic
structure, symplectic algorithms have been regarded as the preferred
geometric numerical integrators, because they conserve the symplectic
structure exactly and their numerical energy error are bounded by
a small number over all time-steps \cite{ruth1983canonical,feng1985,feng1986,kang1989construction,forest1989fourth,channell1990symplectic,yoshida1990construction,candy1991symplectic,SSC94,feng1995collected,yoshida1993recent,marsden2001discrete,GNT,feng2010symplectic}.
Generally speaking, symplectic algorithms are often implicit, such
as general symplectic Runge-Kutta methods, and the roots of the implicit
iterations are usually difficult to search exactly for complicated
vector fields. In order to improve the efficiency and accuracy of
long-term simulations for Hamiltonian systems, explicit symplectic
algorithms are desired \cite{zhang2016explicit}, especially for relativistic
dynamics of charged particles, which describes many important multi-timescale
dynamics, such as runaway electrons in tokamaks \cite{knoepfel1979runaway}.
However, explicit methods for the relativistic system \eqref{eq:nr}
is difficult to find, except for the 1st order symplectic Euler method
\cite{levichev2002symplectic,wang2016lorentz}. Channell suggested
an explicit symplectic method which only applies to the case of magnetostatic
field without electrical field \cite{Rynecomputational}. In this
paper, we propose a method to solve this challenging problem.

In relativity, space-time is a 4-dimensional identity. Space and time
should be treated with equal footing. It is thus natural to take time
$t$ and $p_{0}=-H$ as the fourth conjugate pair, and the Hamiltonian
system is 8-dimensional with the proper time $s$ as the time variable
\cite{goldstein1965classical}. The Hamiltonian system is therefore
defined on the cotangent space of space-time. To simplify the notation,
we take $m=1$, $q=1$ and \textbf{$c=1$}. The new Hamiltonian for
the 8-dimensional Hamiltonian system expressed in terms of $(\mathbf{x},\mathbf{p},t,p_{0})$
is
\begin{equation}
\bar{H}(\mathbf{x},\mathbf{p},t,p_{0})=-\dfrac{1}{2}\left[p_{0}+\phi(\mathbf{x},t)\right]^{2}+\dfrac{1}{2}\left[\mathbf{p}-\mathbf{A}(\mathbf{x},t)\right]^{2}\thinspace,\label{eq:nH}
\end{equation}
where the canonical symplectic structure is extended to $d\mathbf{p}\wedge d\mathbf{x}+dp_{0}\wedge dt$.
The Hamiltonian function $\bar{H}$ should vanish for a real particle,
which is known as the mass-shell condition. We develop explicit symplectic
algorithms of order 2 and 3 for the 8-dimensional system specified
by Eq.\,\eqref{eq:nH} using sum-split and generating function methods.
Sum-split method is deemed as an effective tool to construct explicit
symplectic algorithms for sum-separable Hamiltonians. Recently, He
et. al. have developed explicit non-canonical symplectic algorithm
using sum-split method for non-relativistic charged particle dynamics
\cite{he2015explicit,he2015Hamiltonian,xiao2015explicit}. We also
construct explicit symplectic algorithms for non-relativistic dynamics
of charged particles by combining sum-split technique and generating
functions \cite{zhang2016explicit}. It benefits from that the sub-Hamiltonians
are\emph{ product-separable} in the form of
\begin{equation}
H(\mathbf{Z})=\mathbf{p}_{i}f(\mathbf{x})\thinspace,\label{eq:HPS1}
\end{equation}
where explicit symplectic algorithms with accuracy of order 2 and
3 can be constructed by applying the generating function. In this
paper, we sum-split the new Hamiltonian $\bar{H}$ into seven parts,
three of which can be solved explicitly. The other sub-Hamiltonians
are all product-separable in the form of Eq.\,\eqref{eq:HPS1}, which
admit explicit symplectic algorithms based on the generating functions.
Then explicit canonical symplectic algorithms for relativistic charged
particle dynamics in general time-dependent electromagnetic fields
can be constructed by combining the exact flows and explicit symplectic
sub-algorithms in different manners.

The paper is organized as follows. In Sec. II, we construct explicit
symplectic algorithms of order 2 and 3 for relativistic charged particle
dynamics in time-dependent electromagnetic field based on generating
functions. Numerical examples calculated by the developed explicit
symplectic algorithms are given in Sec. III. Results show that our
algorithms give more accurate secular trajectories compared with non-symplectic
Runge-Kutta methods, and has higher efficiency relative to implicit
symplectic methods.

\section{Explicit symplectic algorithms for relativistic charged particle
dynamics}

\begin{figure}
\includegraphics[scale=0.7]{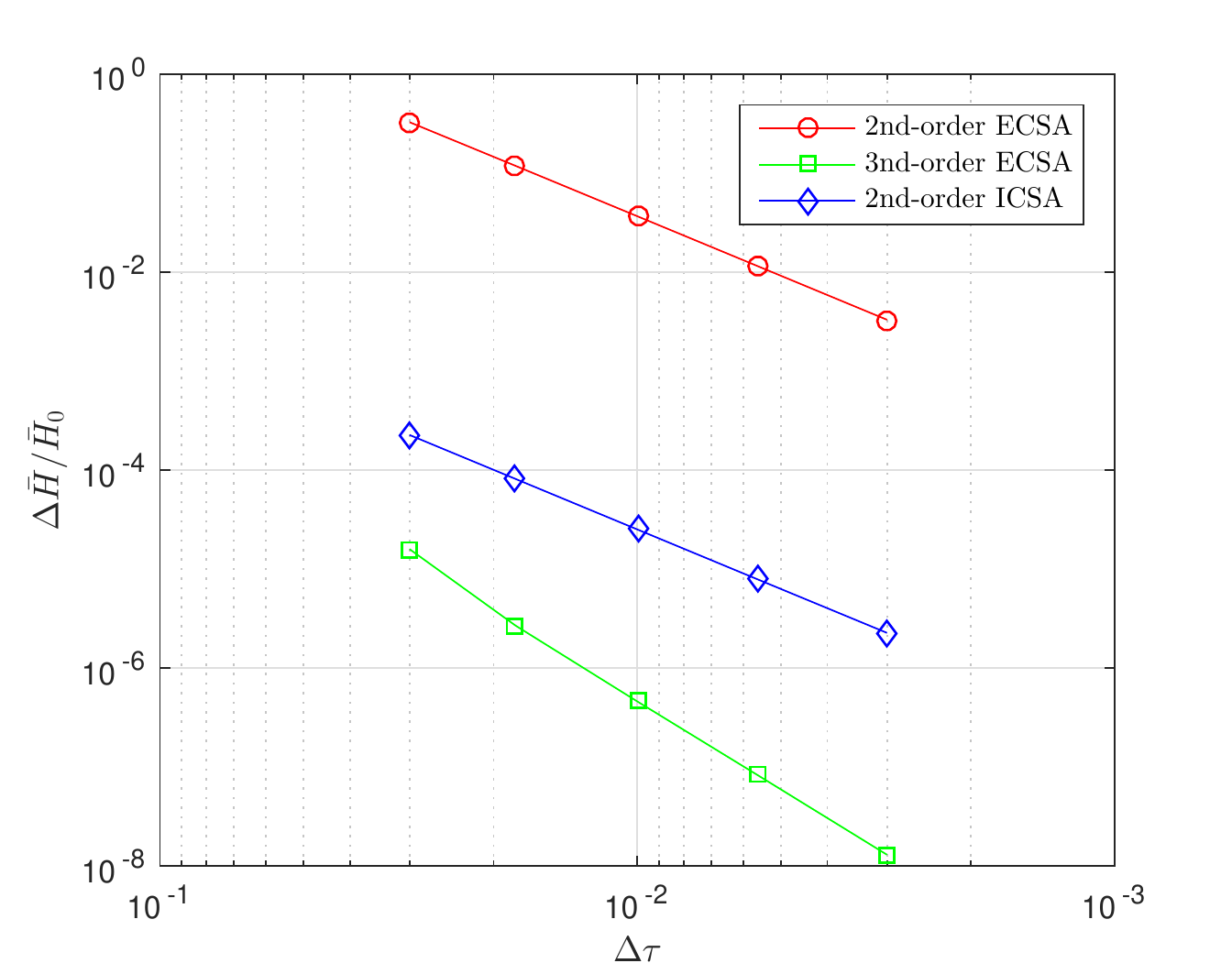}

\protect\caption{Convergence rate of the energy error for three symplectic methods.
It verifies that $\Psi_{\Delta s}^{2}$ is indeed a second order method
and $\Psi_{\Delta s}^{3}$ is a third order method.\label{fig1}}
\end{figure}

The 8-dimensional Hamiltonian system in the extended coordinates $(\mathbf{x},\mathbf{p},t,p_{0})$
is

\begin{equation}
\bar{S}:=\begin{cases}
\dfrac{d\mathbf{x}}{ds} & =\dfrac{\partial\bar{H}}{\partial\mathbf{p}}=\mathbf{p}-\mathbf{A}(\mathbf{x},t)\,,\\
\dfrac{d\mathbf{p}}{ds} & =-\dfrac{\partial\bar{H}}{\partial\mathbf{x}}=\left(\dfrac{\partial\mathbf{A}(\mathbf{x},t)}{\partial\mathbf{x}}\right)^{T}\left[\mathbf{p-}\mathbf{A}(\mathbf{x},t)\right]+\left[p_{0}+\phi(\mathbf{x},t)\right]\nabla\phi\,,\\
\dfrac{dt}{ds} & =\dfrac{\partial\bar{H}}{\partial p_{0}}=-\left[p_{0}+\phi(\mathbf{x},t)\right]\thinspace,\\
\dfrac{dp_{0}}{ds} & =-\dfrac{\partial\bar{H}}{\partial t}=-\dfrac{\partial}{\partial t}\left[-\dfrac{1}{2}\left[p_{0}+\phi(\mathbf{x},t)\right]^{2}+\dfrac{1}{2}\left[\mathbf{p}-\mathbf{A}(\mathbf{x},t)\right]^{2}\right]\thinspace,
\end{cases}
\end{equation}
where $\bar{H}$ is defined by Eq.\,\eqref{eq:nH} and $s$ is the
proper time. For system $\bar{S}$, we sum-split the Hamiltonian function
into seven parts as
\begin{equation}
\bar{H}(\mathbf{x},\mathbf{p},t,p_{0})=H_{1}+H_{2}+H_{3}+H_{4}+H_{5}+H_{6}+H_{7}\,,
\end{equation}
where
\begin{equation}
\begin{split} & H_{1}=\dfrac{1}{2}\mathbf{p}^{2}\,,\qquad H_{2}=\dfrac{1}{2}\mathbf{A}(\mathbf{x},t)^{2}-\dfrac{1}{2}\phi(\mathbf{x},t)^{2}\,,H_{3}=-\dfrac{1}{2}p_{0}^{2}\thinspace,\\
 & H_{4}=-\mathbf{A}(\mathbf{x},t)^{T}\left(p_{1},0,0\right)^{T}=-\mathbf{A}_{1}(\mathbf{x},t)p_{1}\,,\\
 & H_{5}=-\mathbf{A}(\mathbf{x},t)^{T}\left(0,p_{2},0\right)^{T}=-\mathbf{A}_{2}(\mathbf{x},t)p_{2}\,,\\
 & H_{6}=-\mathbf{A}(\mathbf{x},t)^{T}\left(0,0,p_{3}\right)^{T}=-\mathbf{A}_{3}(\mathbf{x},t)p_{3\,.}\\
 & H_{7}=-p_{0}\phi(\mathbf{x},t)\thinspace.
\end{split}
\end{equation}
The corresponding sub-systems generated by these sub-Hamiltonians
are
\begin{alignat}{1}
S_{1}:= & \begin{cases}
\dfrac{d\mathbf{x}}{ds}=\mathbf{p}\,, & \dfrac{d\mathbf{p}}{ds}=\mathbf{0}\,,\\
\dfrac{dt}{ds}=0\thinspace, & \dfrac{dp_{0}}{ds}=0\thinspace,
\end{cases}\label{eq:ss1}\\
S_{2}:= & \begin{cases}
\dfrac{d\mathbf{x}}{ds}=\mathbf{0}\,, & \dfrac{d\mathbf{p}}{ds}=-\left(\dfrac{\partial\mathbf{A}}{\partial\mathbf{x}}\right)^{T}\mathbf{A}(\mathbf{x},t)+\phi(\mathbf{x},t)\nabla\phi\,,\\
\dfrac{dt}{ds}=0\thinspace, & \dfrac{dp_{0}}{ds}=-\dfrac{\partial}{\partial t}\left(\dfrac{1}{2}\mathbf{A}(\mathbf{x},t)^{2}-\dfrac{1}{2}\phi(\mathbf{x},t)^{2}\right)\thinspace,
\end{cases}\label{eq:ss2}\\
S_{3}:= & \begin{cases}
\dfrac{d\mathbf{x}}{dt}=\mathbf{0}\,, & \dfrac{d\mathbf{p}}{dt}=\mathbf{0}\,,\\
\dfrac{dt}{ds}=-p_{0}\thinspace, & \dfrac{dp_{0}}{ds}=0\thinspace,
\end{cases}\\
S_{4}:= & \begin{cases}
\dfrac{d\mathbf{x}}{ds}=\mathbf{-}(\mathbf{A}_{1}(\mathbf{x},t),0,0)^{T}\,, & \dfrac{d\mathbf{p}}{ds}=\left(\dfrac{\partial\mathbf{A}}{\partial\mathbf{x}}(\mathbf{x},t)\right)^{T}\left(p_{1},0,0\right)^{T}\,,\\
\dfrac{dt}{ds}=0\thinspace, & \dfrac{dp_{0}}{ds}=p_{1}\dfrac{\partial\mathbf{A}_{1}}{\partial t}\thinspace,
\end{cases}\label{eq:ss4}\\
S_{5}:= & \begin{cases}
\dfrac{d\mathbf{x}}{dt}=\mathbf{-}(0,\mathbf{A}_{2}(\mathbf{x},t),0)^{T}\,, & \dfrac{d\mathbf{p}}{dt}=\left(\dfrac{\partial\mathbf{A}}{\partial\mathbf{x}}(\mathbf{x},t)\right)^{T}\left(0,p_{2},0\right)^{T}\,,\\
\dfrac{dt}{ds}=0\thinspace, & \dfrac{dp_{0}}{dt}=p_{2}\dfrac{\partial\mathbf{A}_{2}}{\partial t}\thinspace,
\end{cases}\\
S_{6}:= & \begin{cases}
\dfrac{d\mathbf{x}}{dt}=\mathbf{-}(0,0,\mathbf{A}_{3}(\mathbf{x},t),)^{T}\,, & \dfrac{d\mathbf{p}}{dt}=\left(\dfrac{\partial\mathbf{A}}{\partial\mathbf{x}}(\mathbf{x},t)\right)^{T}\left(0,0,p_{3}\right)^{T}\,,\\
\dfrac{dt}{ds}=0\thinspace, & \dfrac{dp_{0}}{ds}=p_{3}\dfrac{\partial\mathbf{A}_{3}}{\partial t}\thinspace,
\end{cases}\\
S_{7}:= & \begin{cases}
\dfrac{d\mathbf{x}}{ds}=\mathbf{0}\thinspace, & \dfrac{d\mathbf{p}}{ds}=p_{0}\nabla\phi\thinspace,\\
\dfrac{dt}{ds}=-\phi(\mathbf{x},t)\thinspace, & \dfrac{dp_{0}}{ds}=p_{0}\dfrac{\partial\phi}{\partial t}\thinspace.
\end{cases}\label{eq:ss7}
\end{alignat}
For subsystems $S_{1}$, $S_{2}$ and $S_{3}$, exact solutions can
be computed explicitly as
\begin{equation}
\begin{alignedat}{1}\varphi_{1}(s):= & \begin{cases}
\mathbf{x}(s)=\mathbf{x}^{0}+s\mathbf{p}^{0}\,, & \mathbf{p}(s)=\mathbf{p}^{0}\,,\\
t(s)=t^{0}\thinspace, & p_{0}(s)=p_{0}^{0}\thinspace,
\end{cases}\\
\varphi_{2}(s):= & \begin{cases}
\mathbf{x}(s)=\mathbf{x}^{0}\,, & \mathbf{p}(s)=\mathbf{p}^{0}-s\left(\dfrac{\partial\mathbf{A}}{\partial\mathbf{x}}\right)^{T}\mathbf{A}(\mathbf{x}^{0},t^{0})+s\phi\nabla\phi(\mathbf{x}^{0},t^{0})\,,\\
t(s)=t^{0}\thinspace, & p_{0}(s)=p_{0}^{0}-s\dfrac{\partial}{\partial t}\left(\dfrac{1}{2}\mathbf{A}(\mathbf{x}^{0},t^{0})^{2}-\dfrac{1}{2}\phi(\mathbf{x}^{0},t^{0})^{2}\right)\thinspace,
\end{cases}\\
\varphi_{3}(s):= & \begin{cases}
\mathbf{x}(s)=\mathbf{x}^{0}\,, & \mathbf{p}(s)=\mathbf{p}^{0}\,,\\
t(s)=t^{0}-sp_{0}^{0}\thinspace, & p_{0}(s)=p_{0}^{0}\thinspace.
\end{cases}
\end{alignedat}
\end{equation}
The Haimiltonian functions of the remaining four subsystems, $S_{4}$,
$S_{5}$, $S_{6}$ and $S_{7}$ are all product-separable in the form
of Eq.\,\eqref{eq:HPS1}, whose explicit symplectic algorithms can
be constructed based on generating functions as described in Ref.\,\cite{zhang2016explicit}.
Let's take the sub-system $S_{4}$ given by Eq.\,\eqref{eq:ss4}
associated with the sub-Hamiltonian $H_{4}(\mathbf{p},\mathbf{x},t,p_{0})=-p_{1}\mathbf{A}_{1}(\mathbf{x},t)$
as an example to demonstrate this method. The symplectic method of
order 2 based on generating function is
\begin{equation}
\begin{alignedat}{1}\mathbf{p}^{n+1} & =\mathbf{p}^{n}-\nabla_{\mathbf{x}}G(\mathbf{p}^{n+1},p_{0}^{n+1},\mathbf{x}^{n},t^{n},\Delta s)\thinspace,\\
\mathbf{x}^{n+1} & =\mathbf{x}^{n}+\nabla_{\mathbf{p}}G(\mathbf{p}^{n+1},p_{0}^{n+1},\mathbf{x}^{n},t^{n},\Delta s)\thinspace,\\
p_{0}^{n+1} & =p_{0}^{n}-\dfrac{\partial G}{\partial t}(\mathbf{p}^{n+1},p_{0}^{n+1},\mathbf{x}^{n},t^{n},\Delta s)\thinspace,\\
t^{n+1} & =t^{n}+\dfrac{\partial G}{\partial p_{0}}(\mathbf{p}^{n+1},p_{0}^{n+1},\mathbf{x}^{n},t^{n},\Delta s)\thinspace,
\end{alignedat}
\label{eq:GF}
\end{equation}
where $G$ is the truncated generating function of order 2,
\begin{equation}
\begin{split}G(\mathbf{p},\mathbf{x},t,p_{0},\Delta s) & =\Delta sH_{4}(\mathbf{p},\mathbf{x},t,p_{0})+\dfrac{\Delta s^{2}}{2}\left(\nabla_{\mathbf{p}}H_{4}\cdot\nabla_{\mathbf{x}}H_{4}\right)(\mathbf{p},\mathbf{x},t,p_{0})\thinspace,\\
 & =-\Delta sp_{1}\mathbf{A}_{1}(\mathbf{x},t)+\dfrac{\Delta s^{2}}{2}p_{1}\dfrac{\partial\mathbf{A}_{1}}{\partial x}\mathbf{A}_{1}(\mathbf{x},t)\thinspace.
\end{split}
\end{equation}
Thus, the second-order explicit symplectic methods $\psi_{4}^{\Delta s}$
for $S_{4}$ is
\begin{equation}
\psi_{4}^{\Delta s}:=\begin{cases}
x^{n+1}=x^{n}-\Delta s\mathbf{A}_{1}(\mathbf{x}^{n},t^{n})+\dfrac{\Delta s^{2}}{2}\left[\mathbf{A}_{1}\dfrac{\partial\mathbf{A}_{1}}{\partial x}\right](\mathbf{x}^{n},t^{n})\,,\\
\mathbf{p}_{1}^{n+1}=\mathbf{p}_{1}^{n}+p_{1}^{n+1}\left[\Delta s\nabla\mathbf{A}_{1}-\dfrac{\Delta s^{2}}{2}\dfrac{\partial\mathbf{A}_{1}}{\partial x}\nabla\mathbf{A}_{1}-\dfrac{\Delta s^{2}}{2}\mathbf{A}_{1}\nabla\dfrac{\partial\mathbf{A}_{1}}{\partial x}\right](\mathbf{x}^{n},t^{n})\,,\\
t^{n+1}=t^{n}\thinspace,\\
p_{0}^{n+1}=p_{0}^{n}+p_{1}^{n+1}\left[\Delta s\dfrac{\partial\mathbf{A}_{1}}{\partial t}-\dfrac{\Delta s^{2}}{2}\dfrac{\partial\mathbf{A}_{1}}{\partial x}\dfrac{\partial\mathbf{A}_{1}}{\partial t}-\dfrac{\Delta s^{2}}{2}\mathbf{A}_{1}\dfrac{\partial^{2}\mathbf{A}_{1}}{\partial x\partial t}\right](\mathbf{x}^{n},t^{n})\thinspace.
\end{cases}
\end{equation}
For sub-systems $S_{5}$, $S_{6}$ and $S_{7}$, second order explicit
symplectic methods $\psi_{5}^{\Delta s}$, $\psi_{6}^{\Delta s}$
and $\psi_{7}^{\Delta s}$ can be constructed using the same method.
Now, we exhibit the symplectic method $\psi_{7}^{\Delta s}$ of order
2 based on similar generating function for the subsystem $S_{7}$
in Eq.\,\eqref{eq:ss7},
\begin{equation}
\psi_{7}^{\Delta s}:=\begin{cases}
x^{n+1}=x^{n}\,,\\
\mathbf{p}^{n+1}=\mathbf{p}^{n}-p_{0}^{n+1}\left[-\Delta s\nabla\phi+\dfrac{\Delta s^{2}}{2}\dfrac{\partial\phi}{\partial t}\nabla\mathbf{A}_{1}+\dfrac{\Delta s^{2}}{2}\phi\nabla\dfrac{\partial\phi}{\partial t}\right](\mathbf{x}^{n},t^{n})\,,\\
t^{n+1}=t^{n}+\left[-\Delta s\phi(\mathbf{x}^{n},t^{n})+\dfrac{\Delta s^{2}}{2}\dfrac{\partial\phi}{\partial t}\phi(\mathbf{x}^{n},t^{n})\right]\thinspace,\\
p_{0}^{n+1}=p_{0}^{n}-p_{0}^{n+1}\left[-\Delta s\dfrac{\partial\phi}{\partial t}+\dfrac{\Delta s^{2}}{2}\dfrac{\partial\phi}{\partial t}\dfrac{\partial\phi}{\partial t}+\dfrac{\Delta s^{2}}{2}\phi\dfrac{\partial^{2}\phi}{\partial t^{2}}\right](\mathbf{x}^{n},t^{n})\thinspace.
\end{cases}
\end{equation}
Composing the exact solutions and the symplectic numerical flows of
the seven subsystems, we obtain the following explicit symplectic
method for charged particle dynamics with the accuracy of order 1,
\begin{equation}
\Psi_{\Delta s}^{1}=\varphi_{1}^{\Delta s}\circ\varphi_{2}^{\Delta s}\circ\varphi_{3}^{\Delta s}\circ\psi_{4}^{\Delta s}\circ\psi_{5}^{\Delta s}\circ\psi_{6}^{\Delta s}\circ\psi_{7}^{\Delta s}\,.
\end{equation}
Combining the sub-flows in the following manner,
\begin{equation}
\Psi_{\Delta s}^{2}=\varphi_{1}^{\Delta s/2}\circ\varphi_{2}^{\Delta s/2}\circ\varphi_{3}^{\Delta s/2}\circ\psi_{4}^{\Delta s/2}\circ\psi_{5}^{\Delta s/2}\circ\psi_{6}^{\Delta s/2}\circ\psi_{7}^{\Delta s}\circ\psi_{6}^{\Delta s/2}\circ\psi_{5}^{\Delta s/2}\circ\psi_{4}^{\Delta s/2}\circ\varphi_{3}^{\Delta s/2}\circ\varphi_{2}^{\Delta s/2}\circ\varphi_{1}^{\Delta s/2}\,,\label{eq:2thes}
\end{equation}
we obtain explicit symplectic algorithm with accuracy of order 2.
Because all the sub-algorithms preserve the canonical symplectic structure,
$\Psi_{\Delta s}^{1}$ and $\Psi_{\Delta s}^{2}$ preserve the canonical
symplectic structure of the extended Hamiltonian system. The accuracy
of $\Psi_{\Delta s}^{2}$ can be calculated using the method given
in Ref.\,\cite{zhang2016explicit}. A third order algorithm can be
obtained by the following composition method using $\Psi_{\Delta s}^{2}$,
\begin{equation}
\Psi_{\Delta s}^{3}=\Psi_{a\Delta s}^{2}\circ\Psi_{b\Delta s}^{2}\circ\Psi_{a\Delta s}^{2}\thinspace,\label{eq:3order}
\end{equation}
where $a=\dfrac{1}{2-2^{1/3}}$ and $b=1-2a$. To verify the accuracy
of the explicit canonical symplectic algorithms (ECSA) $\Psi_{\Delta s}^{2}$
and $\Psi_{\Delta s}^{3}$, we plot the relative errors of Hamiltonian
with respect to the proper time step $\Delta s$ in Fig. \ref{fig1}
by comparing with a second order implicit canonical symplectic algorithm
(ICSA)-the mid-point rule. It is obvious that $\Psi_{\Delta s}^{2}$
has the same order with the mid-ponit rule, which is of order 2, and
$\Psi_{\Delta s}^{3}$ has higher accuracy.

\section{Numerical experiments}

\begin{figure}
\includegraphics[scale=0.7]{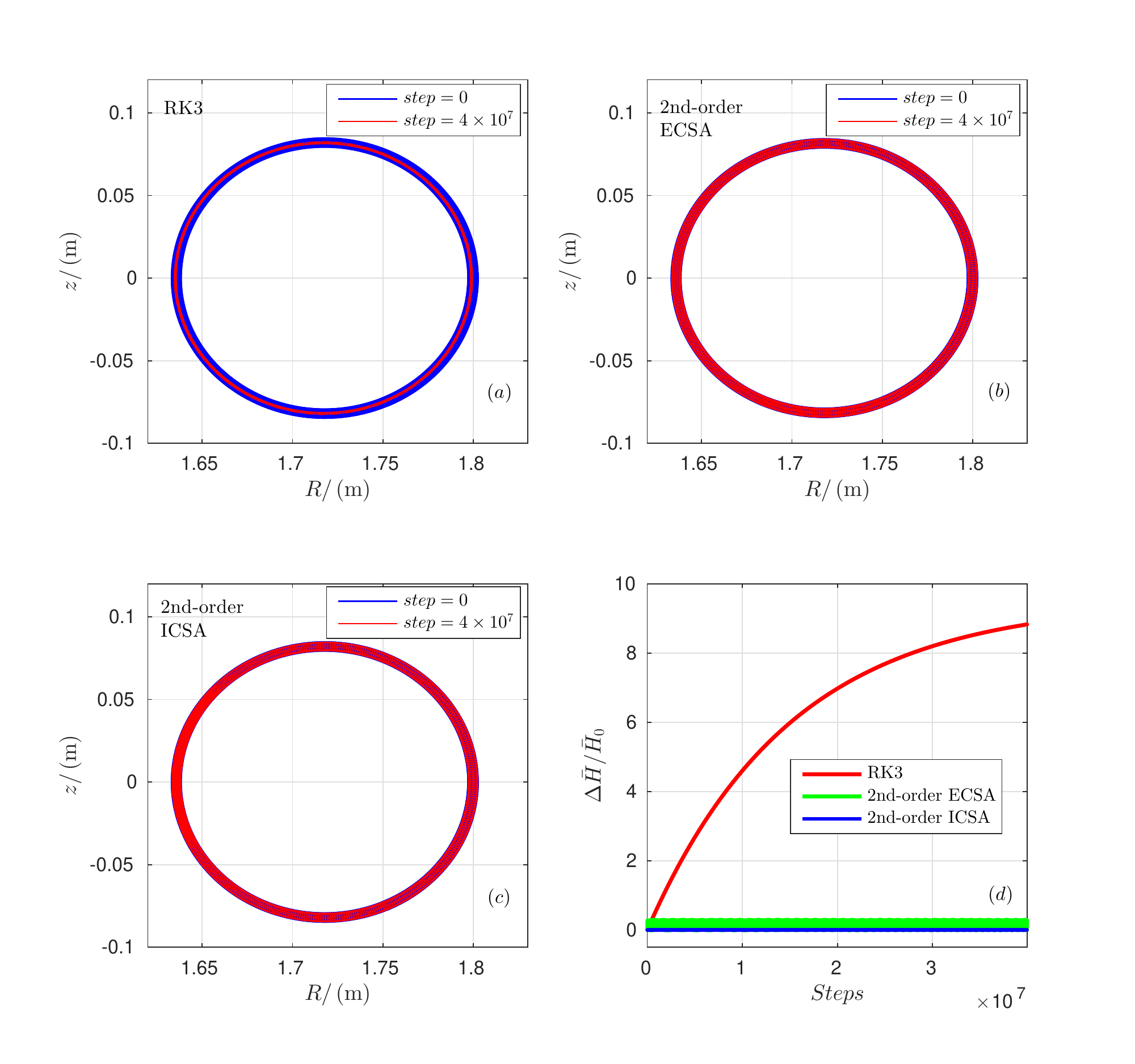}

\protect\caption{Simulations of long-term dynamics of a runaway electron in a tokamak.
The initial orbits are plotted using blue lines, and the orbits after
$4\times10^{7}$ steps are plotted using red lines. (a) Numerical
orbit obtained by a non-symplectic RK3 method. (b) The orbit obtained
by the 2nd-order ECSA method. (c) The numerical orbit by the 2nd-order
ICSA method. (d) The mass-shell error $\Delta\bar{H}/\bar{H}_{0}$
of three methods are plotted as functions of simulation time step.
\label{fig2}}
\end{figure}

To demonstrate the long-term accuracy, conservativeness and efficiency
of developed explicit symplectic algorithms for relativistic charged
particle dynamics, we apply it to the secular runaway dynamics in
tokamak. The electric and magnetic field are chosen to be

\begin{equation}
\begin{split}\mathbf{A}(\mathbf{x},t) & =\mathbf{A}_{0}(\mathbf{x})-\mathbf{E}(\mathbf{x})t\thinspace,\\
\mathbf{B}(\mathbf{x}) & =\nabla\times\mathbf{A}_{0}(\mathbf{x})\thinspace,\\
\mathbf{E}(\mathbf{x}) & =-E_{l}\dfrac{R_{0}}{R}e_{\zeta}\thinspace,\\
\mathbf{A}_{0}(\mathbf{x}) & =\dfrac{B_{0}r^{2}}{2Rq}e_{\zeta}-\ln\left(\dfrac{R}{R_{0}}\right)\dfrac{R_{0}B_{0}}{2}e_{z}+\dfrac{B_{0}R_{0}z}{2R}e_{R}\,.
\end{split}
\end{equation}
where $R=\sqrt{x^{2}+y^{2}}$, $R_{0}$ is the major radius, $B_{0}$
is the magnetic field on axis, the constant $q$ is the safety factor,
and $\zeta=arctan\left(\frac{x}{y}\right)$ is the toroidal coordinate
of the torus. In this example, we take $R_{0}=1.7m$, $E_{l}=2V/m$
and $B_{0}=2T$ with $q=2$. The initial position and momentum of
the runaway electron are $\mathbf{x}_{0}=(1.8,0,0)m$ and $\mathbf{p}_{0}=(3,10,0)m_{0}c$,
where $c$ is the speed of light, and the proper time-step is set
to be $\text{\ensuremath{\Delta}}s=0.03$. Displayed in Fig.$\,$\ref{fig2}
is the comparison of transit orbits calculated by the non-symplectic
third order Runge-Kutta (RK3) method, second order implicit symplectic
mid-point (2nd-order ICSA) method and the explicit second symplectic
(2nd-order ECSA) algorithm $\Psi_{\Delta s}^{2}$. It is expected
that after $4\times10^{7}$ time steps, i.e. $1.136\times10^{-4}$s,
the width of transit orbits is almost the same with that of the initial
orbits, since the diameter of gyromotion expressed by the width of
orbit make little changing. Figure.$\,$\ref{fig2}(a) shows that
the width of orbit obtained by the non-symplectic RK3 method after
$4\times10^{7}$ time steps is smaller than that of the initial one.
The loss of the vertical momentum of runaway electron is mainly due
to the accumulated numerical error of RK3. Meanwhile the orbits calculated
by the 2nd-order ECSA method $\Psi_{\Delta t}^{2}$ in Fig.$\,$\ref{fig2}(b)
and 2nd-order ICSA algorithm in Fig.$\,$\ref{fig2}(c) after $4\times10^{7}$
time steps are almost the same with the initial one. The long-term
relative mass-shell error by non-symplectic method gradually increases
without bound due to numerical errors. On the contrary, for the symplectic
integrators, the relative mass-shell errors are bounded by a small
number for all time. This fact is clearly demonstrated in Fig.$\,$\ref{fig2}(d),
where the mass-shell errors for the three algorithms are plotted.

To illustrate the efficiency of the explicit symplectic algorithms
developed, the CPU time used by the three methods for calculating
the charged particle dynamics is listed in Table.~\ref{Tab1}. The
numerical calculation consists of $4\times10^{6}$ time-steps, and
is carried using on a Inter Core $i5-3210M$ CPU. It's clear that
the 2nd-order ECSA algorithm is much more efficient than the 2nd-order
ICSA algorithm.

\begin{table}
\begin{tabular}{|c|c|c|c|}
\hline
 & RK3 & 2nd-order ICSA & 2nd-order ECSA\tabularnewline
\hline
\hline
CPU time (s) & 6.628 & 29.886 & 14.801\tabularnewline
\hline
\end{tabular}\protect\caption{CPU time used by the three algorithms for runaway dynamics in a tokamak. }
\label{Tab1}
\end{table}

\section{Conclusion}

In this paper, we have constructed explicit symplectic algorithms
for relativistic dynamics of charged particle by extending it into
new variables $(\mathbf{x},\mathbf{p},t,p_{0})$ and combining the
familiar sum-split method with generating function method. Notably,
the developed methodology is expected to be applied to the relativistic
dynamics of charged particle in Yang-Mills fields. In the future,
the explicit symplectic simulation for Vlasov-Maxwell equations of
relativistic charged particles will also be investigated based on
the developed algorithms.
\begin{acknowledgments}
This research is supported by the National Natural Science Foundation
of China (NSFC-11305171, 11505186, 11575185, 11575186), ITER-China
Program (2015GB111003, 2014GB124005), the Fundamental Research Funds
for the Central Universities (No. WK2030040068), China Postdoctoral
Science Foundation (No. 2015M581994), JSPS-NRF-NSFC A3 Foresight Program
(NSFC-11261140328), Key Research Program of Frontier Sciences CAS
(QYZDB-SSW-SYS004) , and the Geo-Algorithmic Plasma Simulator (GAPS)
Project.
\end{acknowledgments}


\begin{thebibliography}{50}%
\makeatletter
\providecommand \@ifxundefined [1]{%
 \@ifx{#1\undefined}
}%
\providecommand \@ifnum [1]{%
 \ifnum #1\expandafter \@firstoftwo
 \else \expandafter \@secondoftwo
 \fi
}%
\providecommand \@ifx [1]{%
 \ifx #1\expandafter \@firstoftwo
 \else \expandafter \@secondoftwo
 \fi
}%
\providecommand \natexlab [1]{#1}%
\providecommand \enquote  [1]{``#1''}%
\providecommand \bibnamefont  [1]{#1}%
\providecommand \bibfnamefont [1]{#1}%
\providecommand \citenamefont [1]{#1}%
\providecommand \href@noop [0]{\@secondoftwo}%
\providecommand \href [0]{\begingroup \@sanitize@url \@href}%
\providecommand \@href[1]{\@@startlink{#1}\@@href}%
\providecommand \@@href[1]{\endgroup#1\@@endlink}%
\providecommand \@sanitize@url [0]{\catcode `\\12\catcode `\$12\catcode
  `\&12\catcode `\#12\catcode `\^12\catcode `\_12\catcode `\%12\relax}%
\providecommand \@@startlink[1]{}%
\providecommand \@@endlink[0]{}%
\providecommand \url  [0]{\begingroup\@sanitize@url \@url }%
\providecommand \@url [1]{\endgroup\@href {#1}{\urlprefix }}%
\providecommand \urlprefix  [0]{URL }%
\providecommand \Eprint [0]{\href }%
\providecommand \doibase [0]{http://dx.doi.org/}%
\providecommand \selectlanguage [0]{\@gobble}%
\providecommand \bibinfo  [0]{\@secondoftwo}%
\providecommand \bibfield  [0]{\@secondoftwo}%
\providecommand \translation [1]{[#1]}%
\providecommand \BibitemOpen [0]{}%
\providecommand \bibitemStop [0]{}%
\providecommand \bibitemNoStop [0]{.\EOS\space}%
\providecommand \EOS [0]{\spacefactor3000\relax}%
\providecommand \BibitemShut  [1]{\csname bibitem#1\endcsname}%
\let\auto@bib@innerbib\@empty
\bibitem [{\citenamefont {Qin}\ and\ \citenamefont
  {Guan}(2008)}]{qin2008variational}%
  \BibitemOpen
  \bibfield  {author} {\bibinfo {author} {\bibfnamefont {H.}~\bibnamefont
  {Qin}}\ and\ \bibinfo {author} {\bibfnamefont {X.}~\bibnamefont {Guan}},\
  }\href@noop {} {\bibfield  {journal} {\bibinfo  {journal} {Physical Review
  Letters}\ }\textbf {\bibinfo {volume} {100}},\ \bibinfo {pages} {035006}
  (\bibinfo {year} {2008})}\BibitemShut {NoStop}%
\bibitem [{\citenamefont {Qin}\ \emph {et~al.}(2009)\citenamefont {Qin},
  \citenamefont {Guan},\ and\ \citenamefont {Tang}}]{qin2009variational}%
  \BibitemOpen
  \bibfield  {author} {\bibinfo {author} {\bibfnamefont {H.}~\bibnamefont
  {Qin}}, \bibinfo {author} {\bibfnamefont {X.}~\bibnamefont {Guan}}, \ and\
  \bibinfo {author} {\bibfnamefont {W.~M.}\ \bibnamefont {Tang}},\ }\href@noop
  {} {\bibfield  {journal} {\bibinfo  {journal} {Physics of Plasmas}\ }\textbf
  {\bibinfo {volume} {16}},\ \bibinfo {pages} {042510} (\bibinfo {year}
  {2009})}\BibitemShut {NoStop}%
\bibitem [{\citenamefont {Guan}\ \emph {et~al.}(2010)\citenamefont {Guan},
  \citenamefont {Qin},\ and\ \citenamefont {Fisch}}]{guan2010phase}%
  \BibitemOpen
  \bibfield  {author} {\bibinfo {author} {\bibfnamefont {X.}~\bibnamefont
  {Guan}}, \bibinfo {author} {\bibfnamefont {H.}~\bibnamefont {Qin}}, \ and\
  \bibinfo {author} {\bibfnamefont {N.~J.}\ \bibnamefont {Fisch}},\ }\href@noop
  {} {\bibfield  {journal} {\bibinfo  {journal} {Physics of Plasmas}\ }\textbf
  {\bibinfo {volume} {17}},\ \bibinfo {pages} {092502} (\bibinfo {year}
  {2010})}\BibitemShut {NoStop}%
\bibitem [{\citenamefont {Squire}\ \emph
  {et~al.}(2012{\natexlab{a}})\citenamefont {Squire}, \citenamefont {Qin},\
  and\ \citenamefont {Tang}}]{squire2012gauge}%
  \BibitemOpen
  \bibfield  {author} {\bibinfo {author} {\bibfnamefont {J.}~\bibnamefont
  {Squire}}, \bibinfo {author} {\bibfnamefont {H.}~\bibnamefont {Qin}}, \ and\
  \bibinfo {author} {\bibfnamefont {W.~M.}\ \bibnamefont {Tang}},\ }\href@noop
  {} {\bibfield  {journal} {\bibinfo  {journal} {Physics of Plasmas}\ }\textbf
  {\bibinfo {volume} {19}},\ \bibinfo {pages} {052501} (\bibinfo {year}
  {2012}{\natexlab{a}})}\BibitemShut {NoStop}%
\bibitem [{\citenamefont {Qin}\ \emph {et~al.}(2013)\citenamefont {Qin},
  \citenamefont {Zhang}, \citenamefont {Xiao}, \citenamefont {Liu},
  \citenamefont {Sun},\ and\ \citenamefont {Tang}}]{qin2013boris}%
  \BibitemOpen
  \bibfield  {author} {\bibinfo {author} {\bibfnamefont {H.}~\bibnamefont
  {Qin}}, \bibinfo {author} {\bibfnamefont {S.}~\bibnamefont {Zhang}}, \bibinfo
  {author} {\bibfnamefont {J.}~\bibnamefont {Xiao}}, \bibinfo {author}
  {\bibfnamefont {J.}~\bibnamefont {Liu}}, \bibinfo {author} {\bibfnamefont
  {Y.}~\bibnamefont {Sun}}, \ and\ \bibinfo {author} {\bibfnamefont {W.~M.}\
  \bibnamefont {Tang}},\ }\href@noop {} {\bibfield  {journal} {\bibinfo
  {journal} {Physics of Plasmas}\ }\textbf {\bibinfo {volume} {20}},\ \bibinfo
  {pages} {084503} (\bibinfo {year} {2013})}\BibitemShut {NoStop}%
\bibitem [{\citenamefont {Liu}\ \emph {et~al.}(2014)\citenamefont {Liu},
  \citenamefont {Qin}, \citenamefont {Fisch}, \citenamefont {Teng},\ and\
  \citenamefont {Wang}}]{liu2014fate}%
  \BibitemOpen
  \bibfield  {author} {\bibinfo {author} {\bibfnamefont {J.}~\bibnamefont
  {Liu}}, \bibinfo {author} {\bibfnamefont {H.}~\bibnamefont {Qin}}, \bibinfo
  {author} {\bibfnamefont {N.~J.}\ \bibnamefont {Fisch}}, \bibinfo {author}
  {\bibfnamefont {Q.}~\bibnamefont {Teng}}, \ and\ \bibinfo {author}
  {\bibfnamefont {X.}~\bibnamefont {Wang}},\ }\href@noop {} {\bibfield
  {journal} {\bibinfo  {journal} {Physics of Plasmas}\ }\textbf {\bibinfo
  {volume} {21}},\ \bibinfo {pages} {064503} (\bibinfo {year}
  {2014})}\BibitemShut {NoStop}%
\bibitem [{\citenamefont {Zhang}\ \emph {et~al.}(2014)\citenamefont {Zhang},
  \citenamefont {Liu}, \citenamefont {Tang}, \citenamefont {Qin}, \citenamefont
  {Xiao},\ and\ \citenamefont {Zhu}}]{zhang2014}%
  \BibitemOpen
  \bibfield  {author} {\bibinfo {author} {\bibfnamefont {R.}~\bibnamefont
  {Zhang}}, \bibinfo {author} {\bibfnamefont {J.}~\bibnamefont {Liu}}, \bibinfo
  {author} {\bibfnamefont {Y.}~\bibnamefont {Tang}}, \bibinfo {author}
  {\bibfnamefont {H.}~\bibnamefont {Qin}}, \bibinfo {author} {\bibfnamefont
  {J.}~\bibnamefont {Xiao}}, \ and\ \bibinfo {author} {\bibfnamefont
  {B.}~\bibnamefont {Zhu}},\ }\href@noop {} {\bibfield  {journal} {\bibinfo
  {journal} {Physics of Plasmas}\ }\textbf {\bibinfo {volume} {21}},\ \bibinfo
  {pages} {032504} (\bibinfo {year} {2014})}\BibitemShut {NoStop}%
\bibitem [{\citenamefont {Zhang}\ \emph {et~al.}(2015)\citenamefont {Zhang},
  \citenamefont {Liu}, \citenamefont {Qin}, \citenamefont {Wang}, \citenamefont
  {He},\ and\ \citenamefont {Sun}}]{zhang2015volume}%
  \BibitemOpen
  \bibfield  {author} {\bibinfo {author} {\bibfnamefont {R.}~\bibnamefont
  {Zhang}}, \bibinfo {author} {\bibfnamefont {J.}~\bibnamefont {Liu}}, \bibinfo
  {author} {\bibfnamefont {H.}~\bibnamefont {Qin}}, \bibinfo {author}
  {\bibfnamefont {Y.}~\bibnamefont {Wang}}, \bibinfo {author} {\bibfnamefont
  {Y.}~\bibnamefont {He}}, \ and\ \bibinfo {author} {\bibfnamefont
  {Y.}~\bibnamefont {Sun}},\ }\href@noop {} {\bibfield  {journal} {\bibinfo
  {journal} {Physics of Plasmas}\ }\textbf {\bibinfo {volume} {22}},\ \bibinfo
  {pages} {044501} (\bibinfo {year} {2015})}\BibitemShut {NoStop}%
\bibitem [{\citenamefont {Zhang}\ \emph
  {et~al.}(2016{\natexlab{a}})\citenamefont {Zhang}, \citenamefont {Liu},
  \citenamefont {Qin}, \citenamefont {Tang}, \citenamefont {He},\ and\
  \citenamefont {Wang}}]{zhangapplication}%
  \BibitemOpen
  \bibfield  {author} {\bibinfo {author} {\bibfnamefont {R.}~\bibnamefont
  {Zhang}}, \bibinfo {author} {\bibfnamefont {J.}~\bibnamefont {Liu}}, \bibinfo
  {author} {\bibfnamefont {H.}~\bibnamefont {Qin}}, \bibinfo {author}
  {\bibfnamefont {Y.}~\bibnamefont {Tang}}, \bibinfo {author} {\bibfnamefont
  {Y.}~\bibnamefont {He}}, \ and\ \bibinfo {author} {\bibfnamefont
  {Y.}~\bibnamefont {Wang}},\ }\href@noop {} {\bibfield  {journal} {\bibinfo
  {journal} {Communications in Computational Physics}\ }\textbf {\bibinfo
  {volume} {19}},\ \bibinfo {pages} {1397} (\bibinfo {year}
  {2016}{\natexlab{a}})}\BibitemShut {NoStop}%
\bibitem [{\citenamefont {Ellison}\ \emph
  {et~al.}(2015{\natexlab{a}})\citenamefont {Ellison}, \citenamefont {Burby},\
  and\ \citenamefont {Qin}}]{ellison2015comment}%
  \BibitemOpen
  \bibfield  {author} {\bibinfo {author} {\bibfnamefont {C.}~\bibnamefont
  {Ellison}}, \bibinfo {author} {\bibfnamefont {J.}~\bibnamefont {Burby}}, \
  and\ \bibinfo {author} {\bibfnamefont {H.}~\bibnamefont {Qin}},\ }\href
  {\doibase http://dx.doi.org/10.1016/j.jcp.2015.09.007} {\bibfield  {journal}
  {\bibinfo  {journal} {Journal of Computational Physics}\ }\textbf {\bibinfo
  {volume} {301}},\ \bibinfo {pages} {489 } (\bibinfo {year}
  {2015}{\natexlab{a}})}\BibitemShut {NoStop}%
\bibitem [{\citenamefont {He}\ \emph {et~al.}(2015{\natexlab{a}})\citenamefont
  {He}, \citenamefont {Sun}, \citenamefont {Liu},\ and\ \citenamefont
  {Qin}}]{he2015volume}%
  \BibitemOpen
  \bibfield  {author} {\bibinfo {author} {\bibfnamefont {Y.}~\bibnamefont
  {He}}, \bibinfo {author} {\bibfnamefont {Y.}~\bibnamefont {Sun}}, \bibinfo
  {author} {\bibfnamefont {J.}~\bibnamefont {Liu}}, \ and\ \bibinfo {author}
  {\bibfnamefont {H.}~\bibnamefont {Qin}},\ }\href@noop {} {\bibfield
  {journal} {\bibinfo  {journal} {Journal of Computational Physics}\ }\textbf
  {\bibinfo {volume} {281}},\ \bibinfo {pages} {135} (\bibinfo {year}
  {2015}{\natexlab{a}})}\BibitemShut {NoStop}%
\bibitem [{\citenamefont {He}\ \emph {et~al.}(2015{\natexlab{b}})\citenamefont
  {He}, \citenamefont {Sun}, \citenamefont {Zhou}, \citenamefont {Liu},\ and\
  \citenamefont {Qin}}]{he2015explicit}%
  \BibitemOpen
  \bibfield  {author} {\bibinfo {author} {\bibfnamefont {Y.}~\bibnamefont
  {He}}, \bibinfo {author} {\bibfnamefont {Y.}~\bibnamefont {Sun}}, \bibinfo
  {author} {\bibfnamefont {Z.}~\bibnamefont {Zhou}}, \bibinfo {author}
  {\bibfnamefont {J.}~\bibnamefont {Liu}}, \ and\ \bibinfo {author}
  {\bibfnamefont {H.}~\bibnamefont {Qin}},\ }\href@noop {} {\bibfield
  {journal} {\bibinfo  {journal} {arXiv preprint arXiv:1509.07794}\ } (\bibinfo
  {year} {2015}{\natexlab{b}})}\BibitemShut {NoStop}%
\bibitem [{\citenamefont {Ellison}\ \emph
  {et~al.}(2015{\natexlab{b}})\citenamefont {Ellison}, \citenamefont {Finn},
  \citenamefont {Qin},\ and\ \citenamefont {Tang}}]{ellison2015development}%
  \BibitemOpen
  \bibfield  {author} {\bibinfo {author} {\bibfnamefont {C.~L.}\ \bibnamefont
  {Ellison}}, \bibinfo {author} {\bibfnamefont {J.}~\bibnamefont {Finn}},
  \bibinfo {author} {\bibfnamefont {H.}~\bibnamefont {Qin}}, \ and\ \bibinfo
  {author} {\bibfnamefont {W.~M.}\ \bibnamefont {Tang}},\ }\href@noop {}
  {\bibfield  {journal} {\bibinfo  {journal} {Plasma Physics and Controlled
  Fusion}\ }\textbf {\bibinfo {volume} {57}},\ \bibinfo {pages} {054007}
  (\bibinfo {year} {2015}{\natexlab{b}})}\BibitemShut {NoStop}%
\bibitem [{\citenamefont {Liu}\ \emph {et~al.}(2015)\citenamefont {Liu},
  \citenamefont {Wang},\ and\ \citenamefont {Qin}}]{liu2015neoclassical}%
  \BibitemOpen
  \bibfield  {author} {\bibinfo {author} {\bibfnamefont {J.}~\bibnamefont
  {Liu}}, \bibinfo {author} {\bibfnamefont {Y.}~\bibnamefont {Wang}}, \ and\
  \bibinfo {author} {\bibfnamefont {H.}~\bibnamefont {Qin}},\ }\href@noop {}
  {\bibfield  {journal} {\bibinfo  {journal} {arXiv preprint arXiv:1510.00780}\
  } (\bibinfo {year} {2015})}\BibitemShut {NoStop}%
\bibitem [{\citenamefont {He}\ \emph {et~al.}(2016)\citenamefont {He},
  \citenamefont {Sun}, \citenamefont {Liu},\ and\ \citenamefont
  {Qin}}]{He16-172}%
  \BibitemOpen
  \bibfield  {author} {\bibinfo {author} {\bibfnamefont {Y.}~\bibnamefont
  {He}}, \bibinfo {author} {\bibfnamefont {Y.}~\bibnamefont {Sun}}, \bibinfo
  {author} {\bibfnamefont {J.}~\bibnamefont {Liu}}, \ and\ \bibinfo {author}
  {\bibfnamefont {H.}~\bibnamefont {Qin}},\ }\href@noop {} {\bibfield
  {journal} {\bibinfo  {journal} {Journal of Computational Physics}\ }\textbf
  {\bibinfo {volume} {305}},\ \bibinfo {pages} {172} (\bibinfo {year}
  {2016})}\BibitemShut {NoStop}%
\bibitem [{\citenamefont {Squire}\ \emph
  {et~al.}(2012{\natexlab{b}})\citenamefont {Squire}, \citenamefont {Qin},\
  and\ \citenamefont {Tang}}]{Squire4748}%
  \BibitemOpen
  \bibfield  {author} {\bibinfo {author} {\bibfnamefont {J.}~\bibnamefont
  {Squire}}, \bibinfo {author} {\bibfnamefont {H.}~\bibnamefont {Qin}}, \ and\
  \bibinfo {author} {\bibfnamefont {W.~M.}\ \bibnamefont {Tang}},\ }\href@noop
  {} {\emph {\bibinfo {title} {Geometric Integration Of The Vlasov-Maxwell
  System With A Variational Particle-in-cell Scheme}}},\ \bibinfo {type} {Tech.
  Rep.}\ \bibinfo {number} {PPPL-4748}\ (\bibinfo  {institution} {Princeton
  Plasma Physics Laboratory},\ \bibinfo {year} {2012})\BibitemShut {NoStop}%
\bibitem [{\citenamefont {Squire}\ \emph
  {et~al.}(2012{\natexlab{c}})\citenamefont {Squire}, \citenamefont {Qin},\
  and\ \citenamefont {Tang}}]{squire2012geometric}%
  \BibitemOpen
  \bibfield  {author} {\bibinfo {author} {\bibfnamefont {J.}~\bibnamefont
  {Squire}}, \bibinfo {author} {\bibfnamefont {H.}~\bibnamefont {Qin}}, \ and\
  \bibinfo {author} {\bibfnamefont {W.~M.}\ \bibnamefont {Tang}},\ }\href@noop
  {} {\bibfield  {journal} {\bibinfo  {journal} {Physics of Plasmas}\ }\textbf
  {\bibinfo {volume} {19}},\ \bibinfo {pages} {084501} (\bibinfo {year}
  {2012}{\natexlab{c}})}\BibitemShut {NoStop}%
\bibitem [{\citenamefont {Xiao}\ \emph {et~al.}(2013)\citenamefont {Xiao},
  \citenamefont {Liu}, \citenamefont {Qin},\ and\ \citenamefont
  {Yu}}]{xiao2013variational}%
  \BibitemOpen
  \bibfield  {author} {\bibinfo {author} {\bibfnamefont {J.}~\bibnamefont
  {Xiao}}, \bibinfo {author} {\bibfnamefont {J.}~\bibnamefont {Liu}}, \bibinfo
  {author} {\bibfnamefont {H.}~\bibnamefont {Qin}}, \ and\ \bibinfo {author}
  {\bibfnamefont {Z.}~\bibnamefont {Yu}},\ }\href@noop {} {\bibfield  {journal}
  {\bibinfo  {journal} {Physics of Plasmas}\ }\textbf {\bibinfo {volume}
  {20}},\ \bibinfo {pages} {102517} (\bibinfo {year} {2013})}\BibitemShut
  {NoStop}%
\bibitem [{\citenamefont {Kraus}(2013)}]{kraus2013variational}%
  \BibitemOpen
  \bibfield  {author} {\bibinfo {author} {\bibfnamefont {M.}~\bibnamefont
  {Kraus}},\ }\href@noop {} {\bibfield  {journal} {\bibinfo  {journal} {arXiv
  preprint arXiv:1307.5665}\ } (\bibinfo {year} {2013})}\BibitemShut {NoStop}%
\bibitem [{\citenamefont {Evstatiev}\ and\ \citenamefont
  {Shadwick}(2013)}]{evstatiev2013variational}%
  \BibitemOpen
  \bibfield  {author} {\bibinfo {author} {\bibfnamefont {E.}~\bibnamefont
  {Evstatiev}}\ and\ \bibinfo {author} {\bibfnamefont {B.}~\bibnamefont
  {Shadwick}},\ }\href@noop {} {\bibfield  {journal} {\bibinfo  {journal}
  {Journal of Computational Physics}\ }\textbf {\bibinfo {volume} {245}},\
  \bibinfo {pages} {376} (\bibinfo {year} {2013})}\BibitemShut {NoStop}%
\bibitem [{\citenamefont {Zhou}\ \emph {et~al.}(2014)\citenamefont {Zhou},
  \citenamefont {Qin}, \citenamefont {Burby},\ and\ \citenamefont
  {Bhattacharjee}}]{zhou2014variational}%
  \BibitemOpen
  \bibfield  {author} {\bibinfo {author} {\bibfnamefont {Y.}~\bibnamefont
  {Zhou}}, \bibinfo {author} {\bibfnamefont {H.}~\bibnamefont {Qin}}, \bibinfo
  {author} {\bibfnamefont {J.}~\bibnamefont {Burby}}, \ and\ \bibinfo {author}
  {\bibfnamefont {A.}~\bibnamefont {Bhattacharjee}},\ }\href@noop {} {\bibfield
   {journal} {\bibinfo  {journal} {Physics of Plasmas (1994-present)}\ }\textbf
  {\bibinfo {volume} {21}},\ \bibinfo {pages} {102109} (\bibinfo {year}
  {2014})}\BibitemShut {NoStop}%
\bibitem [{\citenamefont {Shadwick}\ \emph {et~al.}(2014)\citenamefont
  {Shadwick}, \citenamefont {Stamm},\ and\ \citenamefont
  {Evstatiev}}]{Shadwick14}%
  \BibitemOpen
  \bibfield  {author} {\bibinfo {author} {\bibfnamefont {B.~A.}\ \bibnamefont
  {Shadwick}}, \bibinfo {author} {\bibfnamefont {A.~B.}\ \bibnamefont {Stamm}},
  \ and\ \bibinfo {author} {\bibfnamefont {E.~G.}\ \bibnamefont {Evstatiev}},\
  }\href@noop {} {\bibfield  {journal} {\bibinfo  {journal} {Physics of
  Plasmas}\ }\textbf {\bibinfo {volume} {21}},\ \bibinfo {pages} {055708}
  (\bibinfo {year} {2014})}\BibitemShut {NoStop}%
\bibitem [{\citenamefont {Xiao}\ \emph
  {et~al.}(2015{\natexlab{a}})\citenamefont {Xiao}, \citenamefont {Liu},
  \citenamefont {Qin}, \citenamefont {Yu},\ and\ \citenamefont
  {Xiang}}]{xiao2015variational}%
  \BibitemOpen
  \bibfield  {author} {\bibinfo {author} {\bibfnamefont {J.}~\bibnamefont
  {Xiao}}, \bibinfo {author} {\bibfnamefont {J.}~\bibnamefont {Liu}}, \bibinfo
  {author} {\bibfnamefont {H.}~\bibnamefont {Qin}}, \bibinfo {author}
  {\bibfnamefont {Z.}~\bibnamefont {Yu}}, \ and\ \bibinfo {author}
  {\bibfnamefont {N.}~\bibnamefont {Xiang}},\ }\href@noop {} {\bibfield
  {journal} {\bibinfo  {journal} {Physics of Plasmas (1994-present)}\ }\textbf
  {\bibinfo {volume} {22}},\ \bibinfo {pages} {092305} (\bibinfo {year}
  {2015}{\natexlab{a}})}\BibitemShut {NoStop}%
\bibitem [{\citenamefont {Xiao}\ \emph
  {et~al.}(2015{\natexlab{b}})\citenamefont {Xiao}, \citenamefont {Qin},
  \citenamefont {Liu}, \citenamefont {He}, \citenamefont {Zhang},\ and\
  \citenamefont {Sun}}]{xiao2015explicit}%
  \BibitemOpen
  \bibfield  {author} {\bibinfo {author} {\bibfnamefont {J.}~\bibnamefont
  {Xiao}}, \bibinfo {author} {\bibfnamefont {H.}~\bibnamefont {Qin}}, \bibinfo
  {author} {\bibfnamefont {J.}~\bibnamefont {Liu}}, \bibinfo {author}
  {\bibfnamefont {Y.}~\bibnamefont {He}}, \bibinfo {author} {\bibfnamefont
  {R.}~\bibnamefont {Zhang}}, \ and\ \bibinfo {author} {\bibfnamefont
  {Y.}~\bibnamefont {Sun}},\ }\href@noop {} {\bibfield  {journal} {\bibinfo
  {journal} {Physics of Plasmas}\ }\textbf {\bibinfo {volume} {22}},\ \bibinfo
  {pages} {112504} (\bibinfo {year} {2015}{\natexlab{b}})}\BibitemShut
  {NoStop}%
\bibitem [{\citenamefont {Crouseilles}\ \emph {et~al.}(2015)\citenamefont
  {Crouseilles}, \citenamefont {Einkemmer},\ and\ \citenamefont
  {Faou}}]{crouseilles2015hamiltonian}%
  \BibitemOpen
  \bibfield  {author} {\bibinfo {author} {\bibfnamefont {N.}~\bibnamefont
  {Crouseilles}}, \bibinfo {author} {\bibfnamefont {L.}~\bibnamefont
  {Einkemmer}}, \ and\ \bibinfo {author} {\bibfnamefont {E.}~\bibnamefont
  {Faou}},\ }\href@noop {} {\bibfield  {journal} {\bibinfo  {journal} {Journal
  of Computational Physics}\ }\textbf {\bibinfo {volume} {283}},\ \bibinfo
  {pages} {224} (\bibinfo {year} {2015})}\BibitemShut {NoStop}%
\bibitem [{\citenamefont {Qin}\ \emph {et~al.}(2015)\citenamefont {Qin},
  \citenamefont {He}, \citenamefont {Zhang}, \citenamefont {Liu}, \citenamefont
  {Xiao},\ and\ \citenamefont {Wang}}]{qin2015comment}%
  \BibitemOpen
  \bibfield  {author} {\bibinfo {author} {\bibfnamefont {H.}~\bibnamefont
  {Qin}}, \bibinfo {author} {\bibfnamefont {Y.}~\bibnamefont {He}}, \bibinfo
  {author} {\bibfnamefont {R.}~\bibnamefont {Zhang}}, \bibinfo {author}
  {\bibfnamefont {J.}~\bibnamefont {Liu}}, \bibinfo {author} {\bibfnamefont
  {J.}~\bibnamefont {Xiao}}, \ and\ \bibinfo {author} {\bibfnamefont
  {Y.}~\bibnamefont {Wang}},\ }\href@noop {} {\bibfield  {journal} {\bibinfo
  {journal} {Journal of Computational Physics}\ }\textbf {\bibinfo {volume}
  {297}},\ \bibinfo {pages} {721} (\bibinfo {year} {2015})}\BibitemShut
  {NoStop}%
\bibitem [{\citenamefont {He}\ \emph {et~al.}(2015{\natexlab{c}})\citenamefont
  {He}, \citenamefont {Qin}, \citenamefont {Sun}, \citenamefont {Xiao},
  \citenamefont {Zhang},\ and\ \citenamefont {Liu}}]{he2015Hamiltonian}%
  \BibitemOpen
  \bibfield  {author} {\bibinfo {author} {\bibfnamefont {Y.}~\bibnamefont
  {He}}, \bibinfo {author} {\bibfnamefont {H.}~\bibnamefont {Qin}}, \bibinfo
  {author} {\bibfnamefont {Y.}~\bibnamefont {Sun}}, \bibinfo {author}
  {\bibfnamefont {J.}~\bibnamefont {Xiao}}, \bibinfo {author} {\bibfnamefont
  {R.}~\bibnamefont {Zhang}}, \ and\ \bibinfo {author} {\bibfnamefont
  {J.}~\bibnamefont {Liu}},\ }\href@noop {} {\bibfield  {journal} {\bibinfo
  {journal} {Physics of Plasmas}\ }\textbf {\bibinfo {volume} {22}},\ \bibinfo
  {pages} {124503} (\bibinfo {year} {2015}{\natexlab{c}})}\BibitemShut
  {NoStop}%
\bibitem [{\citenamefont {Qin}\ \emph {et~al.}(2016)\citenamefont {Qin},
  \citenamefont {Liu}, \citenamefont {Xiao}, \citenamefont {Zhang}, ,
  \citenamefont {He}, \citenamefont {Wang}, \citenamefont {Burby},
  \citenamefont {Ellison},\ and\ \citenamefont {Zhou}}]{qin2016canonical}%
  \BibitemOpen
  \bibfield  {author} {\bibinfo {author} {\bibfnamefont {H.}~\bibnamefont
  {Qin}}, \bibinfo {author} {\bibfnamefont {J.}~\bibnamefont {Liu}}, \bibinfo
  {author} {\bibfnamefont {J.}~\bibnamefont {Xiao}}, \bibinfo {author}
  {\bibfnamefont {R.}~\bibnamefont {Zhang}}, , \bibinfo {author} {\bibfnamefont
  {Y.}~\bibnamefont {He}}, \bibinfo {author} {\bibfnamefont {Y.}~\bibnamefont
  {Wang}}, \bibinfo {author} {\bibfnamefont {J.~W.}\ \bibnamefont {Burby}},
  \bibinfo {author} {\bibfnamefont {L.}~\bibnamefont {Ellison}}, \ and\
  \bibinfo {author} {\bibfnamefont {Y.}~\bibnamefont {Zhou}},\ }\href@noop {}
  {\bibfield  {journal} {\bibinfo  {journal} {Nucl. Fusion}\ }\textbf {\bibinfo
  {volume} {56}},\ \bibinfo {pages} {014001} (\bibinfo {year}
  {2016})}\BibitemShut {NoStop}%
\bibitem [{\citenamefont {Zhou}\ \emph {et~al.}(2016)\citenamefont {Zhou},
  \citenamefont {Huang}, \citenamefont {Qin},\ and\ \citenamefont
  {Bhattacharjee}}]{zhou2015formation}%
  \BibitemOpen
  \bibfield  {author} {\bibinfo {author} {\bibfnamefont {Y.}~\bibnamefont
  {Zhou}}, \bibinfo {author} {\bibfnamefont {Y.~M.}\ \bibnamefont {Huang}},
  \bibinfo {author} {\bibfnamefont {H.}~\bibnamefont {Qin}}, \ and\ \bibinfo
  {author} {\bibfnamefont {A.}~\bibnamefont {Bhattacharjee}},\ }\href {\doibase
  10.1103/PhysRevE.93.023205} {\bibfield  {journal} {\bibinfo  {journal} {Phys.
  Rev. E}\ }\textbf {\bibinfo {volume} {93}},\ \bibinfo {pages} {023205}
  (\bibinfo {year} {2016})}\BibitemShut {NoStop}%
\bibitem [{\citenamefont {Webb}(2016)}]{Webb16}%
  \BibitemOpen
  \bibfield  {author} {\bibinfo {author} {\bibfnamefont {S.~D.}\ \bibnamefont
  {Webb}},\ }\href@noop {} {\bibfield  {journal} {\bibinfo  {journal} {Plasma
  Physics and Controlled Fusion}\ }\textbf {\bibinfo {volume} {58}},\ \bibinfo
  {pages} {034007} (\bibinfo {year} {2016})}\BibitemShut {NoStop}%
\bibitem [{\citenamefont {Ruth}(1983)}]{ruth1983canonical}%
  \BibitemOpen
  \bibfield  {author} {\bibinfo {author} {\bibfnamefont {R.~D.}\ \bibnamefont
  {Ruth}},\ }\href@noop {} {\bibfield  {journal} {\bibinfo  {journal} {IEEE
  Trans. Nucl. Sci.}\ }\textbf {\bibinfo {volume} {30}},\ \bibinfo {pages}
  {2669} (\bibinfo {year} {1983})}\BibitemShut {NoStop}%
\bibitem [{\citenamefont {Feng}(1985)}]{feng1985}%
  \BibitemOpen
  \bibfield  {author} {\bibinfo {author} {\bibfnamefont {K.}~\bibnamefont
  {Feng}},\ }in\ \href@noop {} {\emph {\bibinfo {booktitle} {Proceedings of the
  1984 Beijing Symposium on Differential Geometry and Differential
  Equations}}}\ (\bibinfo {organization} {Beijing Science Press},\ \bibinfo
  {year} {1985})\ pp.\ \bibinfo {pages} {42--58}\BibitemShut {NoStop}%
\bibitem [{\citenamefont {Feng}(1986)}]{feng1986}%
  \BibitemOpen
  \bibfield  {author} {\bibinfo {author} {\bibfnamefont {K.}~\bibnamefont
  {Feng}},\ }\href@noop {} {\bibfield  {journal} {\bibinfo  {journal} {Journal
  of Computational Mathematics}\ }\textbf {\bibinfo {volume} {4}},\ \bibinfo
  {pages} {279} (\bibinfo {year} {1986})}\BibitemShut {NoStop}%
\bibitem [{\citenamefont {Kang}\ \emph {et~al.}(1989)\citenamefont {Kang},
  \citenamefont {Wu}, \citenamefont {Qin},\ and\ \citenamefont
  {Wang}}]{kang1989construction}%
  \BibitemOpen
  \bibfield  {author} {\bibinfo {author} {\bibfnamefont {F.}~\bibnamefont
  {Kang}}, \bibinfo {author} {\bibfnamefont {H.~M.}\ \bibnamefont {Wu}},
  \bibinfo {author} {\bibfnamefont {M.~Z.}\ \bibnamefont {Qin}}, \ and\
  \bibinfo {author} {\bibfnamefont {D.~L.}\ \bibnamefont {Wang}},\ }\href@noop
  {} {\bibfield  {journal} {\bibinfo  {journal} {Journal of Computational
  Mathematics}\ }\textbf {\bibinfo {volume} {7}},\ \bibinfo {pages} {71}
  (\bibinfo {year} {1989})}\BibitemShut {NoStop}%
\bibitem [{\citenamefont {Forest}\ and\ \citenamefont
  {Ruth}(1989)}]{forest1989fourth}%
  \BibitemOpen
  \bibfield  {author} {\bibinfo {author} {\bibfnamefont {E.}~\bibnamefont
  {Forest}}\ and\ \bibinfo {author} {\bibfnamefont {R.~D.}\ \bibnamefont
  {Ruth}},\ }\href@noop {} {\bibfield  {journal} {\bibinfo  {journal}
  {Physica}\ }\textbf {\bibinfo {volume} {43}},\ \bibinfo {pages} {105}
  (\bibinfo {year} {1989})}\BibitemShut {NoStop}%
\bibitem [{\citenamefont {Channell}\ and\ \citenamefont
  {Scovel}(1990)}]{channell1990symplectic}%
  \BibitemOpen
  \bibfield  {author} {\bibinfo {author} {\bibfnamefont {P.}~\bibnamefont
  {Channell}}\ and\ \bibinfo {author} {\bibfnamefont {C.}~\bibnamefont
  {Scovel}},\ }\href@noop {} {\bibfield  {journal} {\bibinfo  {journal}
  {Nonlinearity}\ }\textbf {\bibinfo {volume} {3}},\ \bibinfo {pages} {231}
  (\bibinfo {year} {1990})}\BibitemShut {NoStop}%
\bibitem [{\citenamefont {Yoshida}(1990)}]{yoshida1990construction}%
  \BibitemOpen
  \bibfield  {author} {\bibinfo {author} {\bibfnamefont {H.}~\bibnamefont
  {Yoshida}},\ }\href@noop {} {\bibfield  {journal} {\bibinfo  {journal}
  {Physics Letters A}\ }\textbf {\bibinfo {volume} {150}},\ \bibinfo {pages}
  {262} (\bibinfo {year} {1990})}\BibitemShut {NoStop}%
\bibitem [{\citenamefont {Candy}\ and\ \citenamefont
  {Rozmus}(1991)}]{candy1991symplectic}%
  \BibitemOpen
  \bibfield  {author} {\bibinfo {author} {\bibfnamefont {J.}~\bibnamefont
  {Candy}}\ and\ \bibinfo {author} {\bibfnamefont {W.}~\bibnamefont {Rozmus}},\
  }\href@noop {} {\bibfield  {journal} {\bibinfo  {journal} {Journal of
  Computational Physics}\ }\textbf {\bibinfo {volume} {92}},\ \bibinfo {pages}
  {230} (\bibinfo {year} {1991})}\BibitemShut {NoStop}%
\bibitem [{\citenamefont {Sanz-Serna}\ and\ \citenamefont
  {Calvo}(1994)}]{SSC94}%
  \BibitemOpen
  \bibfield  {author} {\bibinfo {author} {\bibfnamefont {J.~M.}\ \bibnamefont
  {Sanz-Serna}}\ and\ \bibinfo {author} {\bibfnamefont {M.~P.}\ \bibnamefont
  {Calvo}},\ }\href@noop {} {\emph {\bibinfo {title} {Numerical hamiltonian
  problems}}},\ Vol.~\bibinfo {volume} {7}\ (\bibinfo  {publisher} {Chapman and
  Hall, London},\ \bibinfo {year} {1994})\BibitemShut {NoStop}%
\bibitem [{\citenamefont {Feng}(1995)}]{feng1995collected}%
  \BibitemOpen
  \bibfield  {author} {\bibinfo {author} {\bibfnamefont {K.}~\bibnamefont
  {Feng}},\ }\href@noop {} {\emph {\bibinfo {title} {Collected works of Feng
  Kang: II}}}\ (\bibinfo {year} {1995})\BibitemShut {NoStop}%
\bibitem [{\citenamefont {Yoshida}(1993)}]{yoshida1993recent}%
  \BibitemOpen
  \bibfield  {author} {\bibinfo {author} {\bibfnamefont {H.}~\bibnamefont
  {Yoshida}},\ }in\ \href@noop {} {\emph {\bibinfo {booktitle} {Qualitative and
  Quantitative Behaviour of Planetary Systems}}}\ (\bibinfo  {publisher}
  {Springer},\ \bibinfo {year} {1993})\ pp.\ \bibinfo {pages}
  {27--43}\BibitemShut {NoStop}%
\bibitem [{\citenamefont {Marsden}\ and\ \citenamefont
  {West}(2001)}]{marsden2001discrete}%
  \BibitemOpen
  \bibfield  {author} {\bibinfo {author} {\bibfnamefont {J.~E.}\ \bibnamefont
  {Marsden}}\ and\ \bibinfo {author} {\bibfnamefont {M.}~\bibnamefont {West}},\
  }\href@noop {} {\bibfield  {journal} {\bibinfo  {journal} {Acta Numerica
  2001}\ }\textbf {\bibinfo {volume} {10}},\ \bibinfo {pages} {357} (\bibinfo
  {year} {2001})}\BibitemShut {NoStop}%
\bibitem [{\citenamefont {Hairer}\ \emph {et~al.}(2006)\citenamefont {Hairer},
  \citenamefont {Lubich},\ and\ \citenamefont {Wanner}}]{GNT}%
  \BibitemOpen
  \bibfield  {author} {\bibinfo {author} {\bibfnamefont {E.}~\bibnamefont
  {Hairer}}, \bibinfo {author} {\bibfnamefont {C.}~\bibnamefont {Lubich}}, \
  and\ \bibinfo {author} {\bibfnamefont {G.}~\bibnamefont {Wanner}},\
  }\href@noop {} {\emph {\bibinfo {title} {Geometric numerical integration:
  structure-preserving algorithms for ordinary differential equations}}},\
  Vol.~\bibinfo {volume} {31}\ (\bibinfo  {publisher} {Springer},\ \bibinfo
  {year} {2006})\BibitemShut {NoStop}%
\bibitem [{\citenamefont {Feng}\ and\ \citenamefont
  {Qin}(2010)}]{feng2010symplectic}%
  \BibitemOpen
  \bibfield  {author} {\bibinfo {author} {\bibfnamefont {K.}~\bibnamefont
  {Feng}}\ and\ \bibinfo {author} {\bibfnamefont {M.}~\bibnamefont {Qin}},\
  }\href@noop {} {\emph {\bibinfo {title} {Symplectic geometric algorithms for
  hamiltonian systems}}}\ (\bibinfo  {publisher} {Springer},\ \bibinfo {year}
  {2010})\BibitemShut {NoStop}%
\bibitem [{\citenamefont {Zhang}\ \emph
  {et~al.}(2016{\natexlab{b}})\citenamefont {Zhang}, \citenamefont {Qin},
  \citenamefont {Tang}, \citenamefont {Liu}, \citenamefont {Yang},\ and\
  \citenamefont {Xiao}}]{zhang2016explicit}%
  \BibitemOpen
  \bibfield  {author} {\bibinfo {author} {\bibfnamefont {R.}~\bibnamefont
  {Zhang}}, \bibinfo {author} {\bibfnamefont {H.}~\bibnamefont {Qin}}, \bibinfo
  {author} {\bibfnamefont {Y.}~\bibnamefont {Tang}}, \bibinfo {author}
  {\bibfnamefont {J.}~\bibnamefont {Liu}}, \bibinfo {author} {\bibfnamefont
  {H.}~\bibnamefont {Yang}}, \ and\ \bibinfo {author} {\bibfnamefont
  {J.}~\bibnamefont {Xiao}},\ }\href@noop {} {\bibfield  {journal} {\bibinfo
  {journal} {Phys.Rev.E}\ }\textbf {\bibinfo {volume} {94}},\ \bibinfo {pages}
  {013205} (\bibinfo {year} {2016}{\natexlab{b}})}\BibitemShut {NoStop}%
\bibitem [{\citenamefont {Knoepfel}\ and\ \citenamefont
  {Spong}(1979)}]{knoepfel1979runaway}%
  \BibitemOpen
  \bibfield  {author} {\bibinfo {author} {\bibfnamefont {H.}~\bibnamefont
  {Knoepfel}}\ and\ \bibinfo {author} {\bibfnamefont {D.}~\bibnamefont
  {Spong}},\ }\href@noop {} {\bibfield  {journal} {\bibinfo  {journal} {Nuclear
  Fusion}\ }\textbf {\bibinfo {volume} {19}},\ \bibinfo {pages} {785} (\bibinfo
  {year} {1979})}\BibitemShut {NoStop}%
\bibitem [{\citenamefont {Levichev}\ and\ \citenamefont
  {Piminov}(2002)}]{levichev2002symplectic}%
  \BibitemOpen
  \bibfield  {author} {\bibinfo {author} {\bibfnamefont {E.}~\bibnamefont
  {Levichev}}\ and\ \bibinfo {author} {\bibfnamefont {P.}~\bibnamefont
  {Piminov}},\ }in\ \href@noop {} {\emph {\bibinfo {booktitle} {Proceedings of
  the EPAC}}}\ (\bibinfo {year} {2002})\ pp.\ \bibinfo {pages}
  {1655--1657}\BibitemShut {NoStop}%
\bibitem [{\citenamefont {Wang}\ \emph {et~al.}(2016)\citenamefont {Wang},
  \citenamefont {Liu},\ and\ \citenamefont {Qin}}]{wang2016lorentz}%
  \BibitemOpen
  \bibfield  {author} {\bibinfo {author} {\bibfnamefont {Y.}~\bibnamefont
  {Wang}}, \bibinfo {author} {\bibfnamefont {J.}~\bibnamefont {Liu}}, \ and\
  \bibinfo {author} {\bibfnamefont {H.}~\bibnamefont {Qin}},\ }\href@noop {}
  {\bibfield  {journal} {\bibinfo  {journal} {arXiv preprint arXiv:1609.07019}\
  } (\bibinfo {year} {2016})}\BibitemShut {NoStop}%
\bibitem [{\citenamefont {Ryne}()}]{Rynecomputational}%
  \BibitemOpen
  \bibfield  {author} {\bibinfo {author} {\bibfnamefont {R.~D.}\ \bibnamefont
  {Ryne}},\ }\href@noop {} {\emph {\bibinfo {title} {Computational Methods in
  Accelerator Physics}}},\ pp.\ \bibinfo {pages} {19--21}\BibitemShut {NoStop}%
\bibitem [{\citenamefont {Goldstein}(1965)}]{goldstein1965classical}%
  \BibitemOpen
  \bibfield  {author} {\bibinfo {author} {\bibfnamefont {H.}~\bibnamefont
  {Goldstein}},\ }\href@noop {} {\emph {\bibinfo {title} {Classical
  mechanics}}}\ (\bibinfo  {publisher} {Pearson Education India},\ \bibinfo
  {year} {1965})\BibitemShut {NoStop}%
\end{thebibliography}

%

\end{document}